# A Random Interaction Forest for Prioritizing Predictive Biomarkers

Zhen Zeng[1*], Yuefeng Lu[2], Judong Shen[1], Wei Zheng[2], Peter Shaw[1], Mary Beth Dorr[1]

[1] MRL, Merck & Co., Inc., Kenilworth, NJ, USA

[2] Sanofi-Aventis US LLC

[*]Corresponding Author: Zhen Zeng (email: zhz43@pitt.edu)

**Abstract**

Precision medicine is becoming a focus in medical research recently, as its implementation brings values to all stakeholders in the healthcare system. Various statistical methodologies have been developed tackling problems in different aspects of this field, e.g., assessing treatment heterogeneity, identifying patient subgroups, or building treatment decision models. However, there is a lack of new tools devoted to selecting and prioritizing predictive biomarkers. We propose a novel tree-based ensemble method, random interaction forest (RIF), to generate predictive importance scores and prioritize candidate biomarkers for constructing refined treatment decision models. RIF was evaluated by comparing with the conventional random forest and univariable regression methods and showed favorable properties under various simulation scenarios. We applied the proposed RIF method to a biomarker dataset from two phase III clinical trials of bezlotoxumab on *Clostridium difficile* infection recurrence and obtained biologically meaningful results.

# Introduction

Precision medicine is a medical model that uses disease subtypes, genetic markers, and other patient-level factors to develop customized treatment with desirable benefit-risk profiles for patients. The notion has been widely discussed in the public and supported by the government [1]. By making optimal treatment decision, the practice of precision medicine helps providers deliver better healthcare to individual patients. From the perspective of public health, targeting treatment primarily to those who benefit helps optimize limited medical resources. In the pharmaceutical industry, the use of predictive biomarkers increases the odds of success in developing new drugs. New treatments that fail on the primary efficacy objective in overall population may show greater efficacy in a subpopulation. Likewise, the occurrence of unacceptable toxicity of new agents in some patients may be linked to specific patient characteristics and therefore this can be avoided if a biomarker can be found to identify patients who are likely to experience the toxicity. In these cases, drugs that may fail in the overall population are given a second chance to bring value to the right patients with the help of predictive biomarker research in the context of precision medicine.

Traditionally, predefined subgroup analysis serves the purpose of precision medicine. The major issue with conventional subgroup analysis is that the selection of the subgroup may be subjective and limited by prior knowledge. An alternative approach is to carry out a series of *post hoc* univariable regression analyses with biomarker-by-treatment interactions. Such an exploratory analysis is more flexible in terms of the inclusion of biomarkers under consideration, but it is not meant to detect complex interaction forms, e.g., three-way interactions or threshold effects. For both methods, multiple comparisons could be a serious concern when there are many predefined subgroups or exploratory biomarkers.

In clinical trials, potential predictive biomarkers, or candidate biomarkers, are generally introduced based on clinical relevance or screening from the hypothesis-generating approaches, such as exploratory pharmacogenetic and pharmacogenomic studies. The number of candidate biomarkers usually ranges from dozens up to a few hundreds. At such a scale, it is not feasible to directly model multiple variables and their interactions parametrically. As a result, data-driven approaches have been extensively exploited. Many tree-based machine learning methods have emerged recently and achieved marked performance from certain aspects, including Interaction Trees [2], Virtual Twins [3], SIDES [4] and SIDEScreen [5], QUINT [6], GUIDE [7], and causal forests [8].

Interaction Trees (IT) is a single-tree based method, splitting by directly searching for the most significant interaction effect. For each possible split value of each biomarker, a linear model is fitted with main effects of the treatment status and the dichotomized biomarker as well as their interaction. Split of a parent node is chosen based on the largest t-statistic corresponding to the interaction term, so that the treatment effect difference between the two child nodes is maximized. IT repeats the split on nodes to construct a large initial tree and then prunes the tree to control for overfitting.

Virtual Twins (VT) has two steps. Let $X$ denote a vector of biomarkers or covariates, $Y$ denote clinical outcomes, and $T$ denote the treatment status, 0 or 1. In the first step, VT predicts treatment outcome differences $\tau$ with $X, T, X \times \{T = 0\}, X \times \{T = 1\}$ using random forest. Assuming the clinical outcome $Y$ is binary with status 0 and 1, the outcome probability differences under the two potential treatments for each patient i is $\tau_i = P(Y_i = 1 | T_i = 1, X_i) - P(Y_i = 1 | T_i = 0, X_i)$. In the second step, a single-tree model is fitted on the treatment outcome differences $\tau$ with $X$.

Although these methods are inspiring, they lack an assessment for variable predictive importance along with many other predictive modeling approaches. Some other methods, e.g., SIDEScreen and GUIDE, do provide importance scores for variable selection. However, their scores either are not dedicated to predictive utility, or have not been compared against and demonstrated superiority over traditional approaches, such as the importance scores from random forests or the ranking based on p-values from univariable regressions.

If we can reasonably trim the candidate biomarkers down to a few highly predictive ones, it will be much easier to build a refined predictive model with a satisfying performance. Simple and straightforward methods, such as CART and regression should work well in such a case. There are also many advanced methods to consider, such as boosting-based methods [9]. Also, in view of practice, the fewer biomarkers involved, the more likely a treatment decision model may be adopted and promoted. Therefore, identification and prioritization of predictive biomarkers is a key step for achieving precision medicine.

Other drawbacks of existing methods include the lack of extensibility to different types of biomarkers and clinical outcomes, e.g., categorical or continuous biomarkers, and binary, continuous, or survival outcomes, as well as issues associated with single-tree models, e.g., local optimum introduced by greediness, inability to model complex biomarker-by-treatment interactions, and overfitting.

To overcome the aforementioned issues, we propose a novel tree-based ensemble method, RIF (Random Interaction Forest), which generates predictive importance scores to prioritize biomarkers and is robust and extensible to various data types of biomarkers and outcomes. We compared the performance of RIF with random forest and univariable regression under various simulation scenarios and also applied it to a real clinical biomarker dataset.

## Methods

The basic idea of the proposed method is to first design a tree model for capturing the biomarker-by-treatment interaction effects, and then aggregate many of such trees to form an ensemble model.

Similar to random forest [10], each tree in RIF is built upon a bootstrap sample. Child nodes are developed by searching for the best split over a subset of features (i.e., biomarkers) randomly selected.

*Construct an individual tree in RIF*

Unlike interaction trees, which directly split on the most significant interaction (i.e., the most significant difference in treatment effects between the two child nodes), trees in the proposed RIF method split on the most significant treatment effect in any of the two child nodes (Figure 1). The steps to build an individual tree can be described as follows.

1) For all potential splits, calculate the p-values for treatment effect in the two child nodes, respectively, denoted as $P_1$ and $P_2$, and let $P_{min} = min\ \{P_1, P_2\}$. The p-value calculation method depends on the outcome type.
2) Choose the split resulting in the smallest $P_{min}$
3) Continue binary split recursively to construct an individual tree
4) Determine the terminal nodes: stop further splitting a node if it will cause the number of patients in any treatment arm in any of the two future child nodes below a pre-specified minimum.

The trees in the proposed method guarantee a large treatment effect in one child node at each split, no matter what happens to the other child node. However, since the splitting is recursive, all terminal nodes will eventually exhibit treatment effects to the maximum extent. Using p-value rather than the absolute treatment effect to guide split helps strike a balance between the node size and effect size. Compared with interaction trees, the splitting in RIF is simpler and more natural for a tree model. Instead of explicitly searching for the maximum interaction effect at each split, interaction effects are intrinsically captured while the trees grow, because splitting based on p-values allows for various treatment effect directions and scales.

The trees in RIF are flexible to model any type of outcomes. It can adapt to continuous outcome by using two-sample t-test for calculating $P_1$ and $P_2$, binary outcome by using chi-square test or exact test, and survival outcome by using log rank test.

To handle all types of biomarkers and reduce the search space for computational efficiency, we first recode categorical biomarkers to ordinal ones according to their marginal treatment effect estimate within each category. In the case that some categories are absent in one treatment arm but present in the other arm and therefore their marginal treatment effects are non-estimable, the median value is assigned to these categories after recoding to minimize their impact on the tree construction. In this way, all biomarkers are recoded to discrete quantiles (the default maximum number of quantiles is set to be 32).

*Prediction of treatment outcomes for new patients*

To predict treatment outcomes for a new patient, we first gather the biomarkers and covariates vector $X$ of the new patient. Base on $X$ and the RIF model trained on the training data, corresponding terminal nodes in all trees can be located. Outcome can be predicted by taking the average outcome values of subjects in the corresponding terminal nodes over all trees for each treatment. Treatment effect is predicted by taking the difference of the predicted outcomes between two treatments. Prediction intervals can be generated using the collected predictions from all trees.

*Importance Score*

The following steps are used to compute variable predictive importance scores:

1) Run RIF to get the prediction model $M$

2) Make predictions of treatment effect $M(D)$ for all the training data $D$

3) Assume the total number of biomarkers is $p$, generate a series of disrupted data $D'_1, ..., D'_p$ by substituting each feature in the training data with its mean, respectively

4) Make predictions of treatment effect for the disrupted data $M(D'_1), ..., M(D'_p)$

5) Define the importance score $IS$ as the squared Euclidean distance between the two sets of predicted treatment effects for all subjects over all trees

$$IS_i = ||M(D) - M(D'_i)||_2^2$$

$$i = 1, ..., p$$

Note that larger important scores indicate higher predictive ability.

## Simulation

### Setup

The performance of RIF on predictive biomarker prioritization was evaluated under four different scenarios by simulation (Table 1). Two treatment arms were considered. In each scenario, 60 candidate biomarkers were simulated, of which 30 are continuous with standard normal distribution and 30 are categorical with 4 levels evenly distributed. Among the 60 candidates, two continuous ones were truly predictive: $X_1$ with both predictive and prognostic effects, and $X_2$ with pure predictive effect. Besides, four biomarkers, $X_3$ to $X_6$, had prognostic effects with various magnitudes. Continuous treatment outcomes were considered in scenarios 1 and 2, with normally distributed residual errors $\varepsilon \sim N(0, 1)$, and binary treatment outcomes were considered in scenarios 3 and 4. In scenario 1 and 3, a linear additive predictive effect was imposed between $X_1$ and $X_2$. In scenario 2 and 4, a joint threshold predictive effect was present between $X_1$ and $X_2$ (i.e., a three-way interaction among $X_1$, $X_2$, and $T$). More details of the underlying true models can be found in Table 1.

Denote the total sample size as $N$. For each scenario, three different sample sizes were tested: a small sample size of $N$=200 (100 in each treatment arm); a medium sample size of $N$=400 (200 in each treatment arm), and a large sample size of $N$=1000 (500 in each treatment arm). 500 simulations were run for each combination of the four scenarios and three sample sizes.

The parameters for random interaction forests were set as: 500 trees in each RIF; $p/3$ candidate biomarkers randomly selected for each split; and $N/100$ minimum number of patients in any treatment arm to determine the terminal nodes.

The proposed method RIF was evaluated by comparing with the two conventional univariable regression and random forest methods. A series of linear regression (for continuous outcomes) or logistic regression (for binary outcomes) were conducted for each biomarker, with treatment and biomarker main effects and their interaction in the model. The p-value from a 2-degrees-of-freedom joint test for the biomarker main effect and interaction was used to rank the predictive importance. We denote this approach as *UnivarLR*. Random forests with default settings [13] were performed on each treatment arm separately to generate two sets of variable importance scores. Biomarkers were ranked based on the better ranking between the two sets. This approach was denoted as *RF*. Alternatively, random forest was performed on the predicted treatment effects from the first step of VT (see introduction), and the variable importance scores were used for ranking, denoted as *RF2*.

*Results*

Figure 2 shows the results of simulation. The curves are analogs to the receiver operating characteristic curve, with y axis being the probability of identifying true predictive variables and x axis being the average number of non-predictive variables included, illustrating the successful rate against the amount of noise allowed across all simulation repeats. The curves were generated based on an increasing number of prioritized variables, starting from one. A curve above the other one indicates the corresponding method outperforms the other method.

No single method dominates all competitors for all scenarios. However, it is noticeable that *RIF* outperforms all the others in scenario 4, where the outcome is binary and the two predictive biomarkers form a joint threshold effect naturally defining a subgroup. With a moderately large sample size ($N$=1000), *RIF* achieved an approximate gain of 0.1 in the probability of identifying the truth compared with the second best, *RF2*. The performance of *RIF* and *RF2* were similar in scenario 3, where the outcome is binary and the predictive effects are linear additive. For scenarios 1 and 2, our method outperformed *RF2* if very few (≤ 3 in most cases) top-ranking biomarkers are in the prioritization list. When more biomarkers are allowed in, *RF2* achieved a better detection power.

In general, the performance of *UnivarLR* and *RF* were worse than *RIF* and *RF2*, and appeared to be better than each other when the true predictive effect was linear additive or a joint threshold, since the former fits the assumption of *UnivarLR* and the latter is easy to be detected by tree models. It is noteworthy that both of these methods target pure prognostic effects, too. On the contrary, *RIF* and *RF2* are robust to the presence of prognostic effects.

**Real Biomarker Data Analysis**

We illustrated our methods on a real biomarker dataset from two phase 3 clinical trials, MODIFY I and MODIFY II [11], investigating the effect of human monoclonal antibodies against *Clostridium difficile* (*C. diff*) toxins on the recurrence of *C. diff* infection (CDI). The primary end point was CDI recurrence, defined as the binary status of the presence of new CDI episode after initial clinical cure, within 12 weeks after infusion of the study medication. We were interested in identifying patient-level variables that influence the efficacy of treatment with bezlotoxumab (human monoclonal antibody against *C. diff* toxin B) versus treatment without bezlotoxumab. A total of 782 eligible subjects who achieved initial clinical cure were included in the analysis. Among the 38 variables considered ($X$), 20 were categorical and 18 were continuous, including patient demographic data, disease status and history, baseline laboratory assessments, host genetic variants and background, pathogen strain type and lineage at baseline, and randomization stratification factors (see supplemental materials for variable annotation). Of note, after excluding redundant significant signals, two host single nucleotide polymorphisms (SNPs) and one human leukocyte antigen (HLA) allele, coded numerically as 0, 1, and 2 (number of allele of interest), as well as five principal components characterizing patient genetic background, from previous pharmacogenetic genome-wide association studies [12], were included in the variable set.

Missing values in $X$ were imputed using the "rfImpute" function from R package "randomForest" [13]. The results of *RIF* were compared with the aforementioned methods in the simulation section. We grew 1000 trees in *RF*, *RF2*, and *RIF*. To determine if a biomarker is significantly predictive with *UnivarLR*, a p-value threshold of 0.0013 was adopted based on Bonferroni correction. For other methods, significance of predictive biomarkers was determined by repeatedly applying Grubbs' test for one outlier [14] on the importance scores. If the variable with the largest important score is significant (p-value < 0.05), it was removed and the same procedure was applied to the rest until all significant variables were identified.

The top 8 most predictive variables prioritized by each method were presented in Table 2. RIF identified two significant variables, a host SNP rs76166871 and REA group. The other host SNP rs2516513 and HLA allele DRB1*07:01 were not significant by Grubbs' test, but were ranked among the top 4. Although the analysis was totally data-driven, these results convey biology explanation and clinical relevance. Note that both SNPs are located on chromosome 6p, where the HLA region is harbored. Along with the HLA allele DRB1*07:01, these host genetic variants may impact the modulation or regulation of immune response to CDI. On the other hand, restriction endonuclease analysis (REA) is a classification of pathogen strain type based on *C. diff* genetic diversity. REA grouping is informative on the toxin classification of the isolates, including toxin variant types. Therefore, *RIF* identified and prioritized host and pathogen genetic factors as the most predictive variables. These findings supported the involvement of host and pathogen genetic factors, as well as a possible interplay, in the mechanism of action of bezlotoxumab.

Other variables in the top lists worth mentioning were principal components (PC1 to PC5, relevant to patient genetic background), country name (CNTRYNAM, a confounder to both principal components and pathogen diversity), age (AGE, a known prognostic variable), status of prior history of CDI in the past 6 months (HX6M), status of any prior history of CDI (CDIEVER), and number of CDI episodes in the past 6 months (EP6MN).

It was not surprising that *UnivarLR* identified all three host genetics variants as significant, as they were from previous pharmacogenetics screening using the same analysis approach. *UnivarLR* also identified the most variables. This may be due to an inadequate power of Grubbs' test when the number of candidate variables is relatively small or when many truly predictive ones exist. Therefore, Grubbs' test should be used with caution. Instead, an arbitrary cut-off may be more practical to select variables in that the bottleneck is the feasibility of building a final treatment decision model.

## Discussion

Our method enjoyed several favorable properties compared with RF2. First, RIF was more sensitive under stringent variable selection. In other words, if our focus is the top few variables (≤ 3 in our simulation cases) in the prioritization list, RIF is more likely to pick up the truth than other methods. This is of important practical value, since it is essential to keep treatment decision models as simple and sparse as possible so that they can be easily tested and implemented in reality. Any decision model requiring more than three biomarkers is almost prohibitive in practice. Secondly, RIF showed a considerable improvement for scenario 4 when the sample size is moderately large, which is a common and reasonable case in reality. Actually, our real data analysis example is one of such cases. However, there is still a lack of theoretical explanation for these simulation findings. Future endeavors on the theoretical analysis to back these properties up are warranted. Thirdly, RF2 may be biased toward categorical variables with many levels, because many dummy variables and interaction terms have to be introduced for handling them, while RIF is not affected by this issue.

Sometimes, it is desirable to adjust for nuisance covariates. For example, there are cases where variables known to impact the outcomes are not considered as candidates for predictive effect. One way to handle this is to first regress out the effects of these variables, and then applying RIF on the outcome residuals. However, this only works for continuous outcomes. For binary or survival outcomes, as a future work, we may consider using a regression model to adjust for nuisance covariates when splitting nodes. The other way is to include nuisance covariates as variables in RIF, but simply ignore them when interpreting the results and selecting predictive variables. In our real data analysis, randomization stratification factors (the standard-of-care and hospitalization status) and principal components were such nuisance variables. The need to adjust for nuisance covariates is more common when dealing with data from observational studies.

Although our method was proposed in the context of biomarker prioritization for the use in precision medicine, it is applicable to any predictive variables analysis and problem of interaction detection. In the present work, we only investigated RIF's performance on predictive variable prioritization. It is warranted to explore its performance on other uses and possible extensions, such as for individual treatment effect prediction, on survival outcomes, on multivariate outcomes, as well as on longitudinal data.

To summarize, our methods provide a powerful tool, RIF, for prioritizing predictive biomarkers. RIF is implemented in R. The code is available for research purposes upon request.

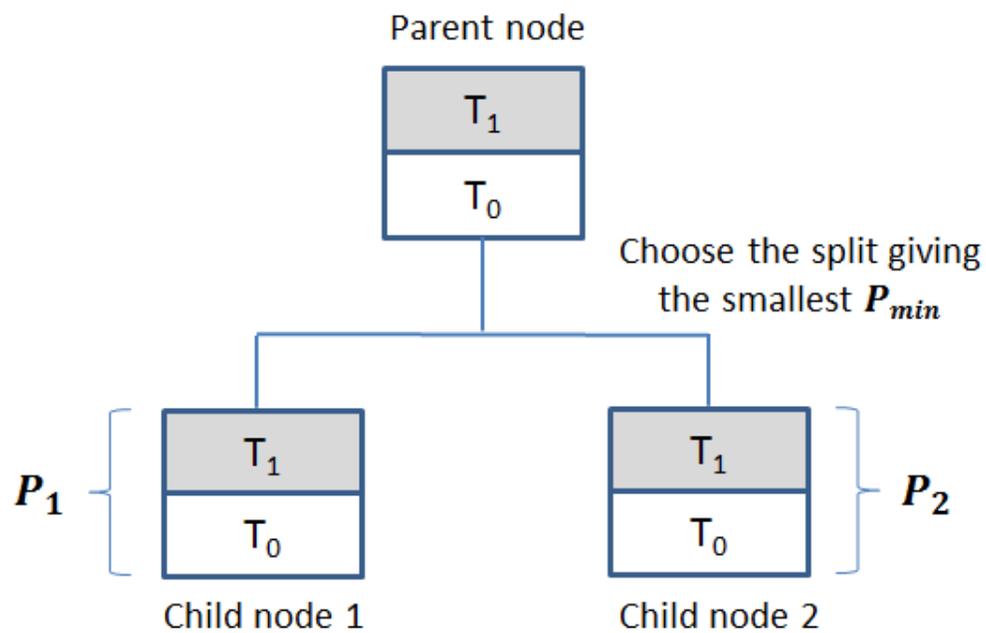

Figure 1. Define the best split of a node to construct trees in RIF.

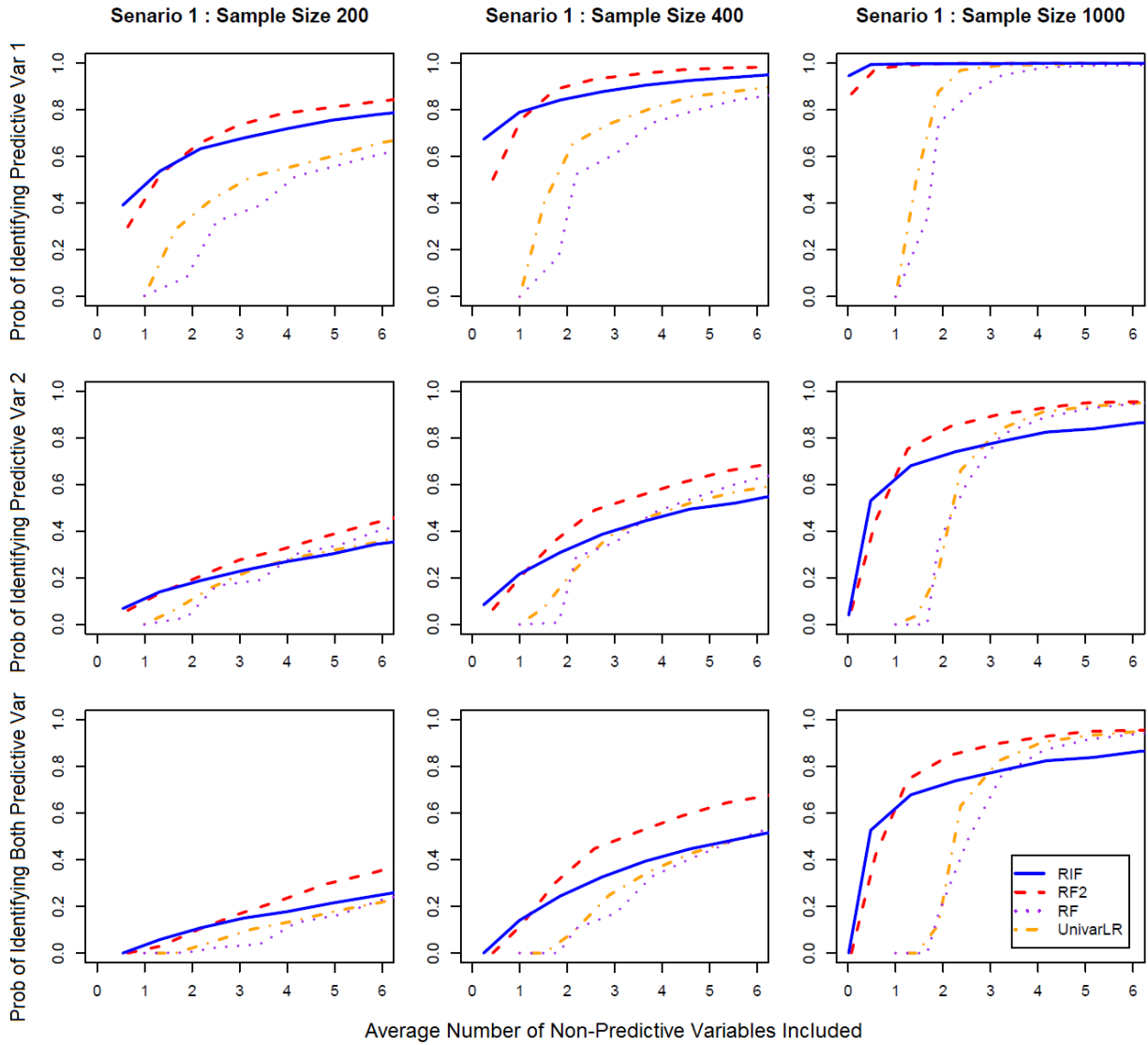

a. Scenario 1

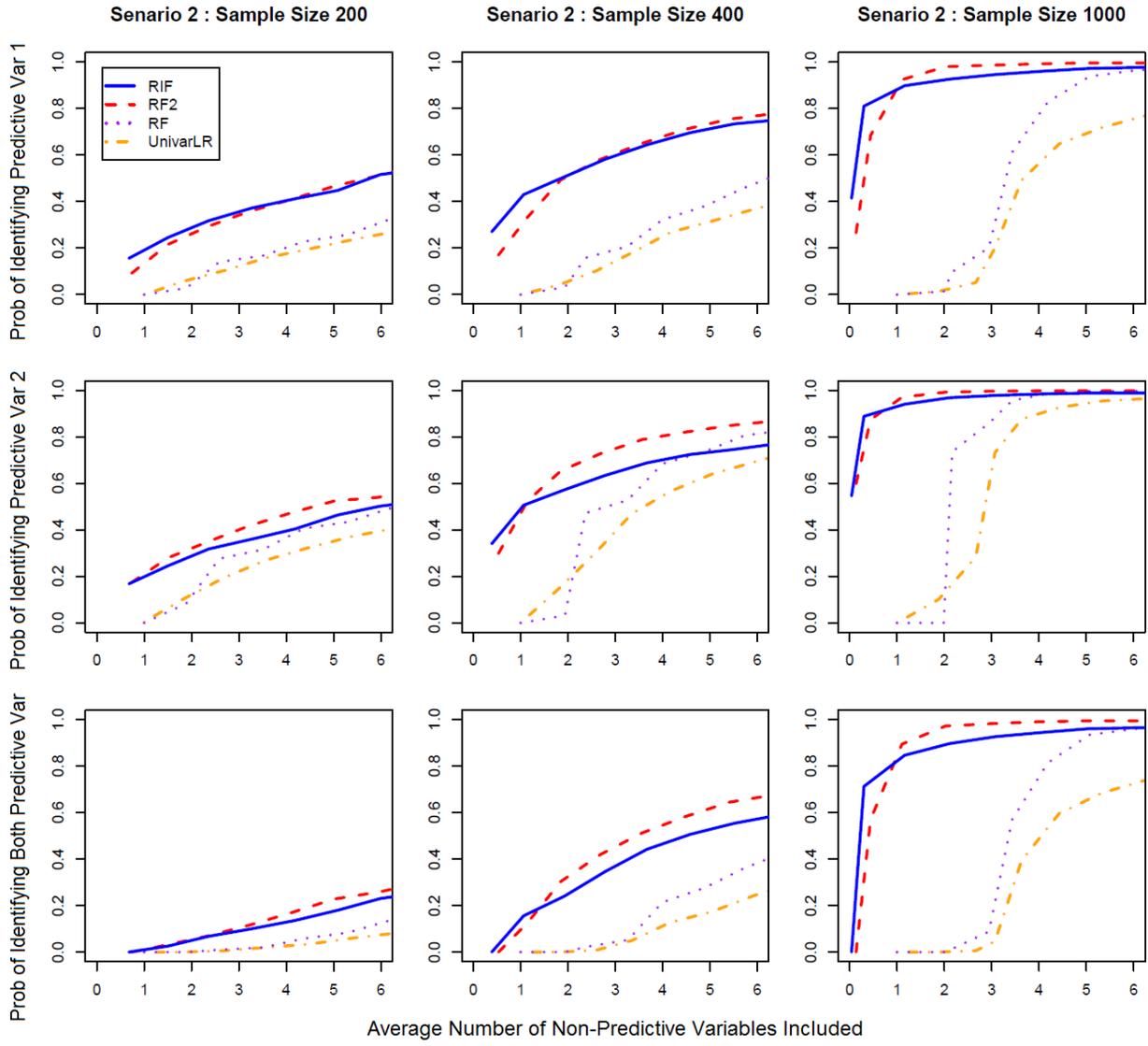

b. Scenario 2

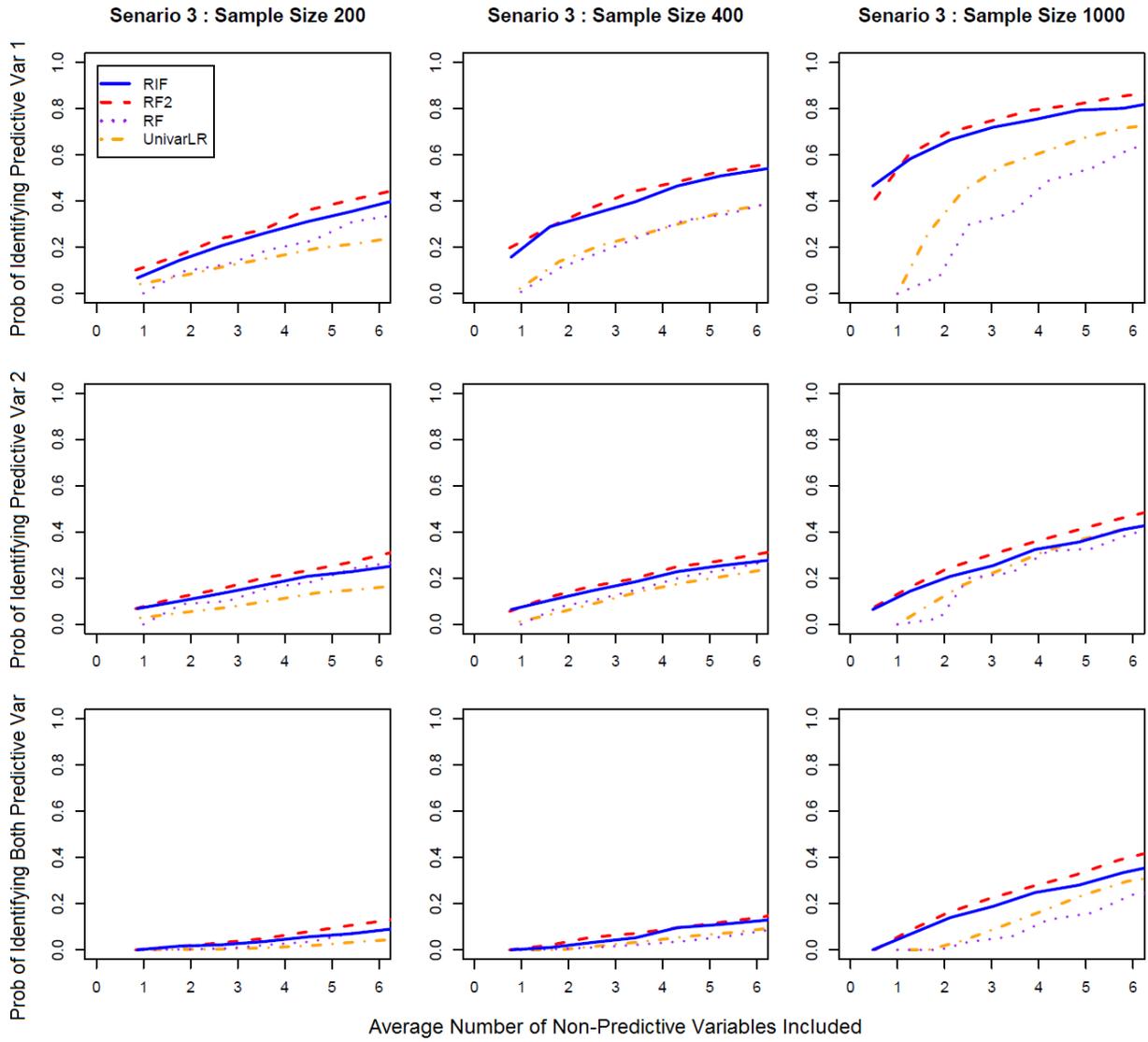

c. Scenario 3

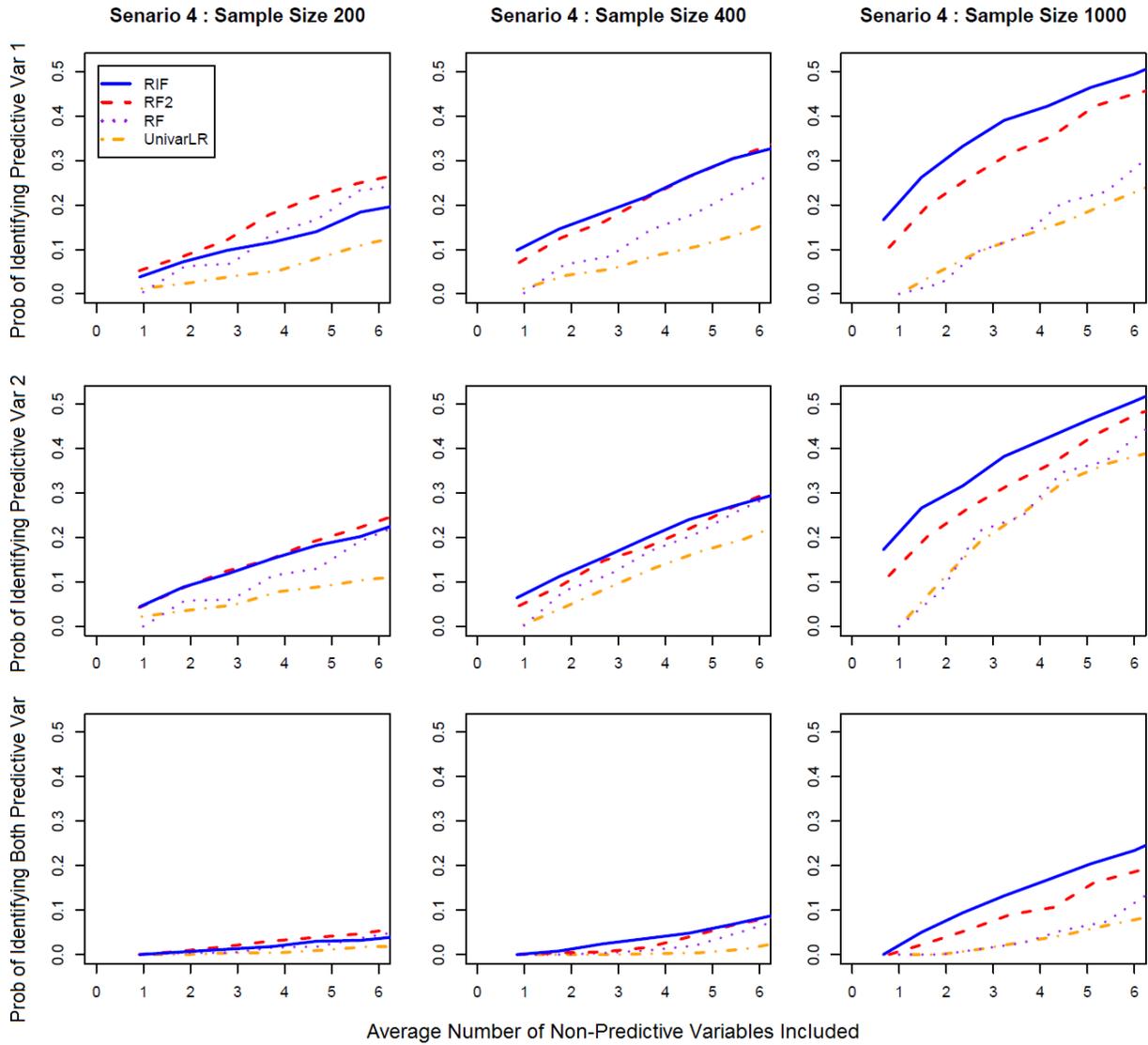

d. Scenario 4

Figure 2. Simulation results of the four scenarios. Each panel represents one scenario. The results of different sample sizes are displayed by column within each panel. Four different methods are indicated by line type. The curves show the probabilities of identifying the true predictive variables (y-axis): $X_1$ (top row in each panel), $X_2$ (middle row in each panel), and both (bottom row in each panel), against the average number of non-predictive variables included (x-axis), with the number of variables in the prioritization list increasing.

Table 1. True models of the four simulation scenarios. Two predictive variables $X_1$ and $X_2$ are highlighted in bold. The visualization of each scenario displays the varying treatment effect as the values of $X_1$ (in x-axis) and $X_2$ (in y-axis) increase. Red, black, and green colors indicate high, medium, and low values of treatment effect, respectively.

| | Scenario | True Model | Visualization |
|---|---|---|---|
| Continuous Outcomes | Scenario 1 | $Y = 1 + 0.25T + (\mathbf{-0.4X_1 + 0.2X_2})T + \mathbf{0.2X_1} + 0.1X_3 + 0.1X_4 + 0.2X_5 + 0.5X_6 + \varepsilon$ | 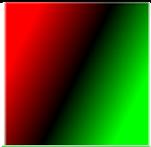 |
| | Scenario 2 | $Y = 1 + 0.25T + \mathbf{I\{X_1 < 0.2, X_2 < 0.2\}}T + \mathbf{0.1X_1} + 0.1X_3 + 0.2X_4 + 0.2X_5 + 0.5X_6 + \varepsilon$ | 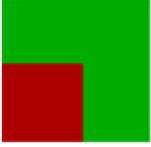 |
| Binary Outcomes | Scenario 3 | $\text{Logit}(E[Y]) = 1 + 0.25T + (\mathbf{-0.4X_1 + 0.2X_2})T + \mathbf{0.2X_1} + 0.1X_3 + 0.1X_4 + 0.2X_5 + 0.5X_6$ | 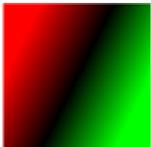 |
| | Scenario 4 | $\text{Logit}(E[Y]) = 1 + 0.25T + \mathbf{I\{X_1 < 0.2, X_2 < 0.2\}}T + \mathbf{0.1X_1} + 0.1X_3 + 0.2X_4 + 0.2X_5 + 0.5X_6$ | 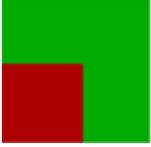 |

Table 2. Top 8 most predictive variables ranked by different methods (see supplemental materials for full results).

| Method | Rank | 1 | 2 | 3 | 4 | 5 | 6 | 7 | 8 |
|---|---|---|---|---|---|---|---|---|---|
| **UnivarLR** | **Variable** | rs76166871 | rs2516513 | DRB1*07:01 | HX6M | CDIEVER | EP6MN | HEPATIC | HYPVGRP |
| | **P-value** | **3.4E-08** | **2.0E-07** | **1.6E-05** | **3.8E-05** | **1.2E-04** | **5.4E-04** | 2.2E-03 | 5.1E-03 |
| | Grubbs' test | NA[b] | NA | NA | NA | NA | NA | NA | NA |
| RF T = 0 | Variable | REA | CNTRYNAM | PC1 | AGE | WEIGHT | PC3 | PC2 | PC5 |
| | IS[a] | 11.15 | 10.63 | 9.16 | 8.73 | 8.34 | 7.89 | 7.71 | 7.46 |
| | Grubbs' test | NS[c] | NS | NS | NS | NS | NS | NS | NS |
| RF T = 1 | Variable | CNTRYNAM | PC4 | REA | ALB_VAL | PC5 | WEIGHT | PC2 | PC1 |
| | IS | 13.81 | 10.57 | 10.35 | 9.80 | 9.68 | 8.96 | 8.88 | 8.53 |
| | Grubbs' test | NS | NS | NS | NS | NS | NS | NS | NS |
| RF2 | Variable | **CNTRYNAM** | REA | rs76166871 | AGE | PC4 | WEIGHT | ALB_VAL | rs2516513 |
| | IS | **6.24** | 5.19 | 3.74 | 3.54 | 3.37 | 3.21 | 3.13 | 3.13 |
| | Grubbs' test | 0.026 | NS | NS | NS | NS | NS | NS | NS |
| RIF | Variable | **rs76166871** | **REA** | rs2516513 | DRB1*07:01 | PC4 | PC2 | AGE | CNTRYNAM |
| | IS | **0.256** | **0.081** | 0.061 | 0.047 | 0.047 | 0.046 | 0.042 | 0.037 |
| | Grubbs' test | 8.9E-13 | 0.015 | NS | NS | NS | NS | NS | NS |

a. IS: important score; b. NA: not available; c. NS: not significant. Significant predictive variables were highlighted in bold.